DRAFT MAY 2018

# The Dataset Nutrition Label:
# A Framework To Drive Higher Data Quality Standards

Sarah Holland[1]\*, Ahmed Hosny[2]\*, Sarah Newman[3], Joshua Joseph[4], and Kasia Chmielinski[1]\*†

[1]*Assembly, MIT Media Lab and Berkman Klein Center at Harvard University,* [2]*Dana-Farber Cancer Institute, Harvard Medical School,* [3]*metaLAB (at) Harvard, Berkman Klein Center for Internet & Society, Harvard University,* [4]*33x.ai*
\**authors contributed equally*
†*nutrition@media.mit.edu*

**ABSTRACT**

Artificial intelligence (AI) systems built on incomplete or biased data will often exhibit problematic outcomes. Current methods of data analysis, particularly before model development, are costly and not standardized. The Dataset Nutrition Label[1] (the Label) is a diagnostic framework that lowers the barrier to standardized data analysis by providing a distilled yet comprehensive overview of dataset "ingredients" before AI model development. Building a Label that can be applied across domains and data types requires that the framework itself be flexible and adaptable; as such, the Label is comprised of diverse qualitative and quantitative modules generated through multiple statistical and probabilistic modelling backends, but displayed in a standardized format. To demonstrate and advance this concept, we generated and published an open source prototype[2] with seven sample modules on the ProPublica Dollars for Docs dataset. The benefits of the Label are manyfold. For data specialists, the Label will drive more robust data analysis practices, provide an efficient way to select the best dataset for their purposes, and increase the overall quality of AI models as a result of more robust training datasets and the ability to check for issues at the time of model development. For those building and publishing datasets, the Label creates an expectation of explanation, which will drive better data collection practices. We also explore the limitations of the Label, including the challenges of generalizing across diverse datasets, and the risk of using "ground truth" data as a comparison dataset. We discuss ways to move forward given the limitations identified. Lastly, we lay out future directions for the Dataset Nutrition Label project, including research and public policy agendas to further advance consideration of the concept.

---

[1] http://datanutrition.media.mit.edu/
[2] http://datanutrition.media.mit.edu/demo.html

## 1   INTRODUCTION

Data driven decision making systems play an increasingly important and impactful role in our lives. These frameworks are built on increasingly sophisticated artificial intelligence (AI) systems and are tuned by a growing population of data specialists[3] to infer a vast diversity of outcomes: the song that plays next on your playlist, the type of advertisement you are most likely to see, or whether you qualify for a mortgage and at what rate [1]. These systems deliver untold societal and economic benefits, but they can also pose harm. Researchers continue to uncover troubling consequences of these systems [2,3].

Data is a fundamental ingredient in AI, and the quality of a dataset used to build a model will directly influence the outcomes it produces. Like the fruit of a poisoned tree, an AI model trained on problematic or missing data will likely produce problematic outcomes [4, 5]. Examples of these problems include gender bias in language translations surfaced through natural language processing [4], and skin shade bias in facial recognition systems due to non-representative data [5]. Typically the model development pipeline (**Figure 1**) begins with a question or goal. Within the realm of supervised learning, for example, a data specialist will curate a labeled dataset of previous answers in response to the guiding question. Such data is then used to train a model to respond in a way that accurately correlates with past occurrences. In this way, past answers are used to forecast the future. This is particularly problematic when outcomes of past events are contaminated with (often unintentional) bias.

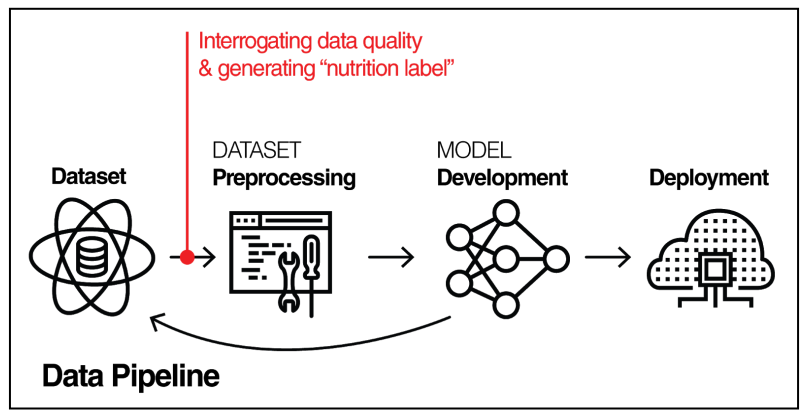

**Figure 1.** Model Development Pipeline

Models often come under scrutiny only after they are built, trained, and deployed. If a model is found to perpetuate a bias - for example, over-indexing for a particular race or gender - the data specialist returns to the development stage in order to identify and address the issue. This feedback loop is inefficient, costly, and does not always mitigate harm; the time and energy of the data specialist is a sunk cost, and if in use, the model may have already caused harm. Some of this harm could be avoided by undertaking thorough interrogation of data at the outset of model development. However, this is still not a widespread or standardized practice.

We conducted an anonymous online survey **(Figure 2),** the results of which further lend credence to this problem. Although many (47%) respondents report conducting some form of data analysis prior to model development, most (74%) indicate that their organizations do not have explicit best practices for such analysis. Fifty-nine percent of respondents reported relying primarily on experience and self-directed learning (through online tutorials, blogs, academic papers, stack overflow, and online data competitions) to inform their data analysis methods and practices. This survey indicates that despite limited current standards, there is widespread interest to improve data analysis practices and make them accessible and standardized.

---

[3] The term "data specialist" is used instead of "data scientist" in the interest of using a term that is broadly scoped to include all professionals utilizing data in automated decision making systems: data scientists, analysts, machine learning engineers, model developers, artificial intelligence researchers, and a variety of others in this space.

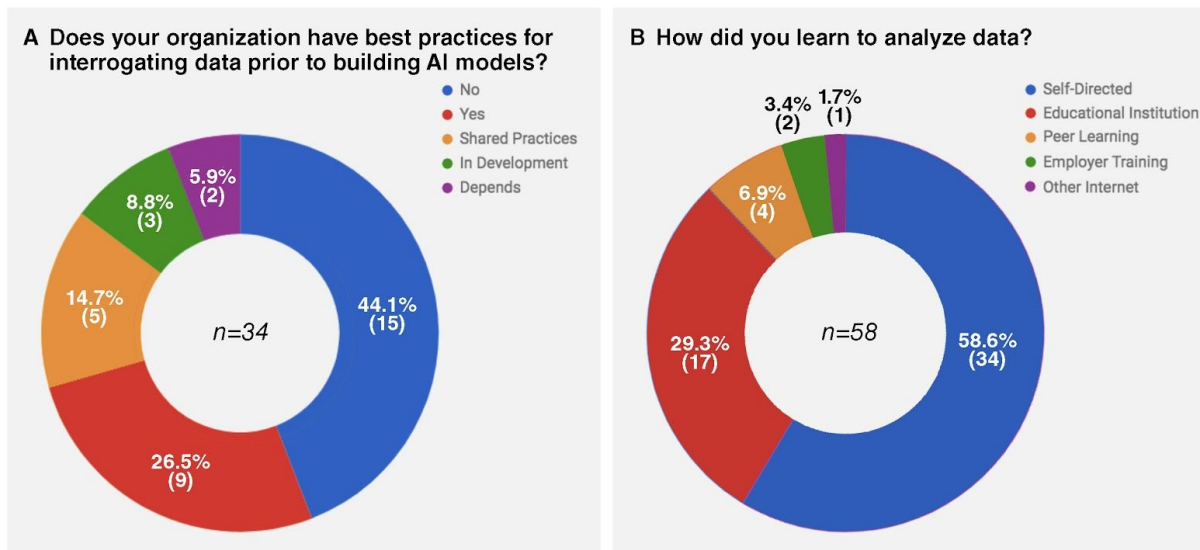

**Figure 2.** (A) Survey results about data analysis best practices in respondents' organizations and (B) Survey results about how respondents learned to analyze data

To improve the accuracy and fairness of AI systems, it is imperative that data specialists are able to more quickly assess the viability and fitness of datasets, and more easily find and use better quality data to train their models. As a proposed solution, we introduce the Dataset Nutrition Label, a diagnostic framework to address and mitigate some of these challenges by providing critical information to data specialists at the point of data analysis.

This study begins with a review of related work, drawing from the fields of nutrition and privacy, where labels are a useful mechanism to distill essential information and enable better decision-making and influence best practices. We then discuss the Dataset Nutrition Label prototype and our methodology, demonstration dataset, and key results. This is followed by an overview of the benefits of the tool, its potential limitations, and ways to mitigate those limitations. We briefly summarize some future directions, including research and public policy agendas that would further advance the goals of the Label. Lastly, we discuss implementation of the prototype and key takeaways.

## 1.1 LABELS IN CONTEXT

To inform the development of our prototype and concept, we surveyed the literature for labeling efforts. Labels and warnings are utilized effectively in product safety [6], pharmaceuticals [7], energy [8], and material safety [9]. We largely draw from the fields of nutrition, online privacy, and algorithmic accountability as they are particularly salient for our purposes. The former is the canonical example and a long standing practice subject to significant study; the latter provides valuable insights in the application of a "nutrition label" in other domains, particularly in subjective contexts and where there is an absence of legal mandates. Collectively, they elucidate the impacts of labels on audience engagement, education, and user decision making.

In 1990, Congress passed the Nutrition Labeling and Education Act (P.L. 101 - 535), which includes a requirement that certain foodstuffs display a standardized "Nutrition Facts" label [10]. By mandating the label, vital nutritional facts were communicated in the context of the "Daily Value" benchmark, and consumers could quickly assess nutrition information and more effectively abide by dietary recommendations at the moment of decision [10–12]. In the nearly three decades since its implementation, several studies have examined the efficacy of the now ubiquitous "Nutrition Facts" label; these studies include analyses of how consumers use the label [11,13], and the effect it has had on the market [14].

Though some cast doubt on the benefits of the mandate in light of its costs [15], most research concludes that the "Nutrition Facts" label has positive impact [16,17]. Surveys demonstrate widespread consumer awareness of the label, and its influence in decision making around food, despite a relatively short time since the passage of the Nutrition Labeling and Education Act [18]. According to the International Food Information Council, more than 80% of consumers reported they looked at the "Nutrition Facts" label when deciding what foods to purchase or consume, and only four percent reported never using the label [19]. Five years after the mandate, the Food Marketing Institute found that about one-third of consumers stopped buying a food because of what they read on the label [18]. With regard to the information contained on the label and consumer understanding, researchers found that "label format and inclusion of (external) reference value information appear to have (positive) effects on consumer perceptions and evaluations," [20] but consumers indicated confusion about the "Daily Value" comparison, suggesting that more information about the source and reliability of ground truth information would be useful [19]. The literature focuses primarily on the impact to consumers rather than on industry operations. However, the significant impact of reported sales and marketing materials on consumers [14] provides a foundation for further inquiry into how this has affected the greater food industry.

In the field of privacy and privacy disclosures, the nutrition label serves as a useful point of reference and inspiration [21]. Researchers at Carnegie Mellon and Microsoft created the "Privacy Nutrition Label" to better surface essential privacy information to assist consumer decision making with regard to the collection, use, and sharing of personal information [22]. The "Privacy Nutrition Label" operates much like "Nutrition Facts" and sits atop existing disclosures. It improves the functionality of the Platform for Privacy Notices, a machine readable format developed by the World Wide Web Consortium, itself an effort to standardize and improve legibility of privacy policies [23]. User surveys that tested the "Privacy Nutrition Label" against alternative formats found that the label outperformed alternatives with "significant positive effects on the accuracy and speed of information finding and reader enjoyment with privacy policies," as well as improved consumer understanding [22,23].

Ranking and scoring algorithms also pose challenges in terms of their complexity, opacity, and sensitivity to the influence of data. End users and even model developers face difficulty in interpreting an algorithm and its ranking outputs, and this difficulty is further compounded when the model and the data on which it is trained is proprietary or otherwise confidential, as is often the case. "Ranking Facts" is a web-based system that generates a "nutrition label" for scoring and ranking algorithms based on factors or "widgets" to communicate an algorithm's methodology or output [24]. Here, the label serves more as an interpretability tool than as a summary of information as the "Nutrition Facts" and "Privacy Nutrition Label," operate. The widgets work together, not modularly, to assess the algorithm on transparency, fairness, stability, and diversity. The demonstration scenarios for using real datasets from college rankings, criminal risk assessment, and financial services establish that the label can be applied to a diverse range of domains. This lends credence to the potential utility in other fields as well, including the rapidly evolving field of AI.

## 1.2 RELATED WORK

More recently, in an effort to improve transparency, accountability, and outcomes of AI systems, AI researchers have proposed methods for standardizing practices and communicating information about the data itself.

The first draws from computer hardware and industry safety standards where datasheets are an industry-wide standard. In datasets, however, they are a novel concept. Datasheets are functionally comparable to the label concept and, like labels that by and large objectively surface empirical information, can often include other information such as recommended uses which are more subjective. "Datasheets for Datasets," a proposal from researchers at Microsoft Research, Georgia Tech, University of Maryland, and the AI Now Institute seeks to standardize information about public datasets, commercial APIs, and pretrained models. The proposed datasheet includes dataset provenance, key characteristics, relevant regulations and test results, but also significant yet more subjective information such as potential bias, strengths and weaknesses of the dataset, API, or model, and suggested uses [25]. As domain experts, dataset, API, and model creators would be responsible for creating the datasheets, not end users or other parties.

We are also aware of a forthcoming study from the field of natural language processing (NLP), "Data Statements for NLP: Toward Mitigating System Bias and Enabling Better Science" [26]. The researchers seek to address ethics, exclusion, and bias issues in NLP systems. Borrowing from similar practices in other fields of practice, the position paper puts forward the concept and practice of "data statements," which are qualitative summaries that provide detailed information and important context about the populations the datasets represent. The information contained in data statements can be used to surface potential mismatches between the populations used to train a system and the populations in planned use prior to deployment, to help diagnose sources of bias that are discovered in deployed systems, and to help understand how experimental results might generalize. The paper's authors suggest that data statements should eventually become required practice for system documentation and academic publications for NLP systems and should be extended to other data types (e.g. image data) albeit with tailored schema.

We take a different, yet complementary, approach. We hypothesize that the concept of a "nutrition label" for datasets is an effective means to provide a scalable and efficient tool to improve the process of dataset interrogation and analysis prior to and during model development. In supporting our hypothesis, we created a prototype, the Dataset Nutrition Label (the Label). Three goals drive this work. First, to inform and improve data specialists' selection and interrogation of datasets and to prompt critical analysis. Consequently, data specialists are the primary intended audience. Second, to gain traction as a practical, readily deployable tool, we prioritize efficiency and flexibility. To that end, we do not suggest one specific approach to the Label, or charge one specific community with creating the Label. Rather, our prototype is modular, and the underlying framework is one that anyone can use. Lastly, we leverage probabilistic computing tools to surface potential corollaries, anomalies, and proxies. This is particularly beneficial because resolving these issues requires excess development time, and can lead to undesired correlations in trained models.

## 2      METHODS

Some assumptions are made to focus our prototyping efforts. Only tabular data is considered. Additionally, we limit our explorations to datasets <10k rows. This allows for a narrower scope and deeper analysis. The Label's first contribution lies in the standard format it provides for metadata communication. This works to address weaknesses in the most common format for tabular data exchange: comma separated values, or the ".csv" format. Despite its simple plain-text nature, portability, and interoperability [27], the lack of additional .csv metadata describing how data should be interpreted, validated, and displayed, is perhaps its biggest drawback. As early as 2015, the World Wide Web Consortium published recommendations on "Metadata Vocabulary for Tabular Data" and "Access methods for CSV Metadata" [28,29]. However, the adoption of these recommendations within the data science community is not widespread. The Label also builds on existing data science practices: directly following the acquisition of a dataset, most data specialists often enter an "exploratory phase". This can be seen, for instance, on web-hosted machine learning competition platforms such as Kaggle, and involves understanding dataset distributions through histograms and other basic statistics. The Label attempts to provide these statistics "out of the box," with the hopes of shortening model development lead times. The Label also aims to provide insights from advanced probabilistic modelling backends for more advanced users. While targeted mainly at a professional audience, many of the modules do not require expert training for interpretation and can thus be utilized in a public-facing Dataset Nutrition Label.

### 2.1      MODULAR ARCHITECTURE

The Label is designed in an extensible fashion with multiple distinct components that we refer to as "modules" (**Table 1**). The modules are stand-alone, allowing for greater flexibility as arrangements of different modules can be used for different types of datasets. This format also caters to a wide range of requirements and information available for a specific dataset. During label generation and subsequent updates, it also accommodates data specialists of different backgrounds and technical skill levels.

Modules (**Table 1 & 2**) range from the purely non-technical, such as the Metadata module, to the highly technical, such as the Probabilistic Computing module. Some modules require manual effort to generate, such as those that provide qualitative descriptions of the data (Metadata, Provenance, Variables), while others can ideally be the result of an automated process (Statistics, Pair Plots). Modules also vary in their subjectivity, especially where there exists a reliance on the Label author to identify which questions should be asked of the data and in what way (e.g. Probabilistic Computing). Many of the example modules are also interactive, highlighting a crucial benefit of a label living on a platform (such as a web page) that supports user interaction. This allows Label users to interrogate various dataset aspects with great flexibility and free of preconceived notions developed during Label generation. Lastly, some modules could be designed to act as proxies for their corresponding dataset as they do not expose the underlying data. This could be key when dealing with proprietary datasets, as much of this data will not or cannot be released to the public based on intellectual property or other constraints. Other modules expose information such as distribution metrics which, in theory, would allow adversaries to approximate the dataset contents. The choice of module(s) is thus based on the availability of information, level of willingness and effort volunteered to document the dataset, and privacy concerns.

| Module Name | Description | Contents |
|---|---|---|
| **Metadata** | Meta information. This module is the only required module. It represents the absolute minimum information to be presented | Filename, file format, URL, domain, keywords, type, dataset size, % of missing cells, license, release date, collection range, description |
| **Provenance** | Information regarding the origin and lineage of the dataset | Source and author contact information with version history |
| **Variables** | Descriptions of each variable (column) in the dataset | Textual descriptions |
| **Statistics** | Simple statistics for all variables, in addition to stratifications into ordinal, nominal, continuous, and discrete | Least/most frequent entries, min/max, median, mean,.etc |
| **Pair Plots** | Distributions and linear correlations between 2 chosen variables | Histograms and heatmaps |
| **Probabilistic Model** | Synthetic data generated using distribution hypotheses from which the data was drawn - leverages a probabilistic programming backend | Histograms and other statistical plots |
| **Ground Truth Correlations** | Linear correlations between a chosen variable in the dataset and variables from other datasets considered to be "ground truth", such as Census Data | Heatmaps |

**Table 1.** Table illustrating 7 modules of the Dataset Nutrition Label, together with their description, role, and contents.

The list of modules currently examined in this study, while not exhaustive, provides a solid representation of the kinds of flexibility supported by the Label framework. Other modules considered for future iterations or additional datasets include but are not limited to: a comments section for users to interact with authors of the Label for feedback or other purposes; an extension of the Provenance section that includes the versioning history and change logs of the dataset and associated Labels over time, similar to Git; a privacy-focused module that indicates any sensitive information and whether the data was collected with consent; and finally, a usage tracking module that documents data utilization and references using some form of identifier, similar to the Digital Object Identifier [30] and associated citation systems in scientific publishing.

| | Module Characteristic - Level Required | | | | |
|---|---|---|---|---|---|
| **Module Name** | Technical Expertise | Manual Effort | Subjectivity | Interactivity | Data Exposure |
| Metadata | Low | High | Low | Low | Low |
| Provenance | Low | High | Low | Low | Low |
| Variables | Low | High | Medium | Low | Medium |
| Statistics | Medium | Low | Low | Low | Medium |
| Pair Plots | Medium | Low | Low | High | High |
| Probabilistic Modeling | High | Medium | High | Low | High |
| Ground Truth Correlations | Medium | Medium | Low | Low | High |

**Table 2.** Variability of attributes across prototype modules highlights the potential diversity of information included in a Label

### 2.2 WEB-BASED APPLICATION

The label is envisioned as a digital object that can be both generated and viewed by web-based applications. The label ecosystem comprises two main components: a label maker and a label viewer (**Figure 3**). Given a specific dataset, the label maker application allows users to select the desired modules and generate them. While the generation of some modules is fully automated, some require human input (**Table 2**). For instance, the Metadata module mainly requires explicit input, while the Pair Plots module can be generated automatically from the dataset. The Label generator pre-populates as many fields as possible and alerts users to those requiring action. The Label itself lives in a .json format, as one that is human readable and well supported. The Label can then be viewed within the label viewer application where formating is carried out to achieve the desired user interface and user interaction effects. In terms of visual appearance and design, format and typeface requirements of the "Nutrition Facts" label [31] is used. These guidelines, such as the all black font color on white contrasting background, are optimized for clarity and conciseness. Design changes are anticipated in further iterations, and should be informed by user testing.

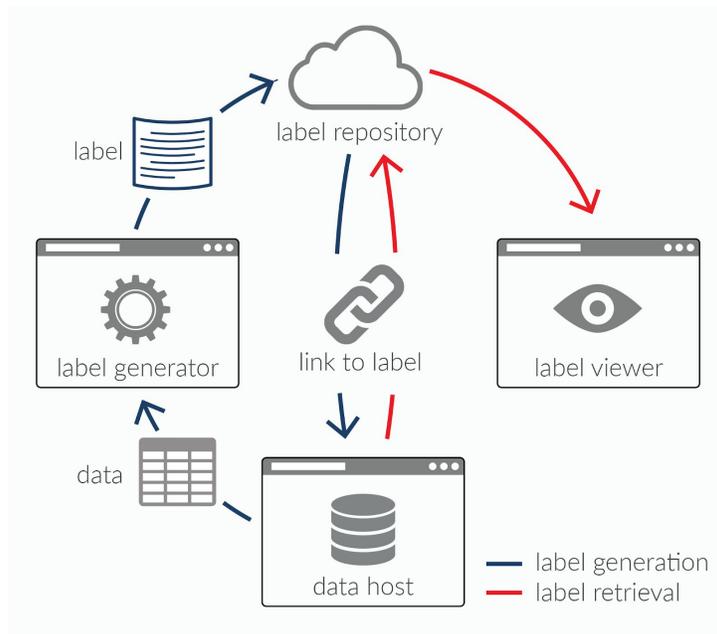

**Figure 3.** Architecture of the proposed Data Nutrition Label ecosystem.

## 2.3     BACKENDS

Simple statistical analyses involving the generation of histograms, distribution information, and linear correlations are carried out directly in the browser, given tabular datasets of <100K rows. Server-side processing is thus reserved for more specialized and sophisticated analyses requiring additional computational power. Such processing could run multiple backends with the ultimate aim of providing the Label authors with a diverse set of options, fueled by the plethora of tools developed by research groups for automating the generation of summaries, insights, and understandings of datasets. The Label thus becomes a medium for the continuous deployment and testing of these tools. A somewhat recent and particularly powerful example of this is probabilistic computing, and specifically, BayesDB [32], an open source platform developed by researchers at MIT. With minimal modeling and programming effort, BayesDB enables inference of a model that captures the structure underlying the data and generates statistical summaries based on such structure.

## 3     RESULTS

To test the concept generally and the modular framework specifically, we built a prototype with a dataset that included information about people and was maintained by an organization invested in better understanding the data. This combination of factors provides necessary information and access to build a wide variety of modules, including those that require full knowledge of the data and the ability to contact the organization that maintains the dataset. We were granted access to the "Dollars for Docs" database from ProPublica, an independent, nonprofit newsroom that produces investigative journalism in the public interest[4]. The dataset, which contains payments to doctors and teaching hospitals from pharmaceutical and medical device companies over a two-year time period (August 2013 - December 2015), was originally released by the U.S. Centers for Medicare and Medicaid Services (CMS) and compiled by ProPublica into a single, comprehensive database.

---

[4] https://projects.propublica.org/docdollars/

The resulting prototype successfully demonstrates how disparate modules can be built on a specific dataset in order to highlight multiple, complementary facets of the data, ideally to be leveraged for further investigation by data specialists through the use of additional tools and strategies. The prototype Label includes seven modules (**Table 1, 2**). The Metadata, Provenance, and Variables modules (**Supp. Figure 1**) provide as-is dataset information. They mirror information submitted by the Label authors as well as provide a standard format for both the generation and consumption of such data. The Statistics module (**Supp. Figure 2**) starts to offer a glimpse into the dataset distributions. For instance, the skewness of a 500 row dataset subset towards a particular drug "Xarelto" can be quickly identified as the most frequent entry under the variable "product_name", and "Aciphex" as the least frequent entry. The Pair Plot module (**Figure 4**) starts to introduce interactivity into the label where the viewer is able to choose the variable pair being compared to one another. A specialist building a model predicting marketing spend in each state, for example, may choose to compare "recipient_state" and "total_amount_of_payment_usdollars," and will observe that some states (CA, NY) are more highly correlated with spend. In this case, the specialist would probably normalize for population as the next step beyond consulting the Label in order to identify anomalous spending trends.

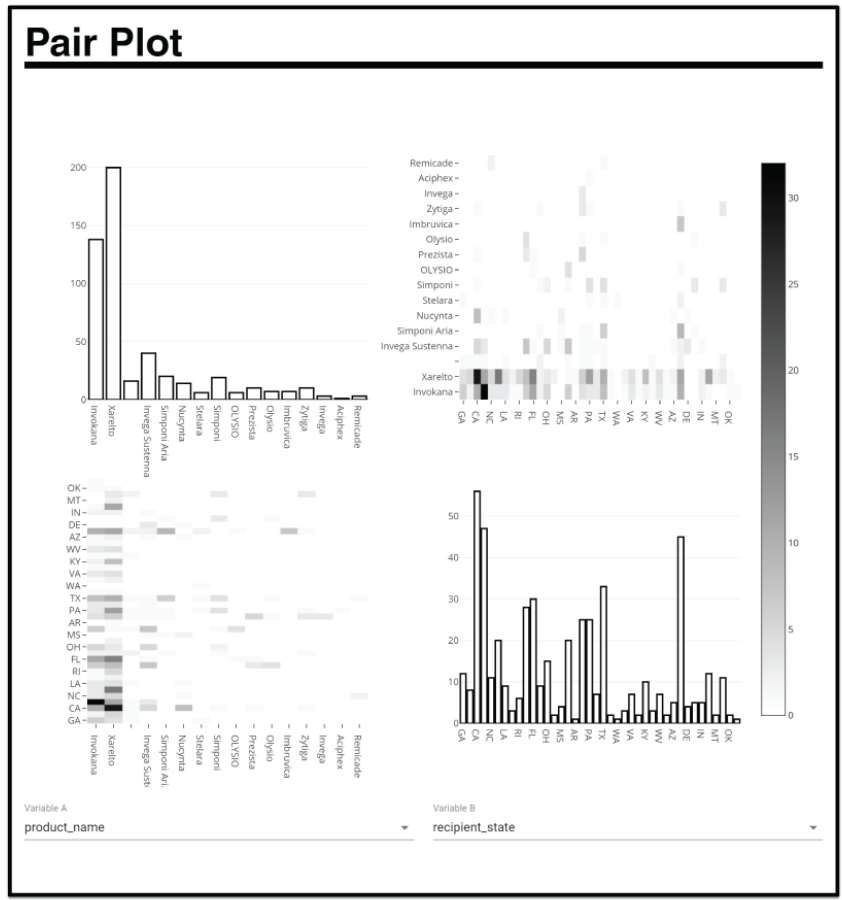

**Figure 4.** Prototype Label demonstrating the Pair Plot module and highlighting the interactive dropdown menus for selecting variables.

While all modules thus far investigate the dataset itself, the Probabilistic Model module (**Figure 5**) attempts to generate synthetic data by utilizing the aforementioned BayesDB backend. Computed from an inferred generative model, this module allows for the full benefits of Bayesian analysis [33], such as interpretability of inferences, coping with missing data, and robustness to outliers and regions of sparse data. In this specific use case, an underrepresented drug is chosen from the dataset and the probability of this drug receiving a payment in different states is inferred. With the inevitable variation in the representation of different groups in datasets, such analyses are of great utility in extracting insights - even from relatively small sample sizes. A quick toggle indicates that the top few states for marketing spend are likely the same few states - with a few exceptions, including that NJ is likely to receive much more money for marketing activities relating to the drug Xarelto. Again, this information only acts as a flag for the "what"; specialists will ideally continue to investigate the data in order to identify the "why".

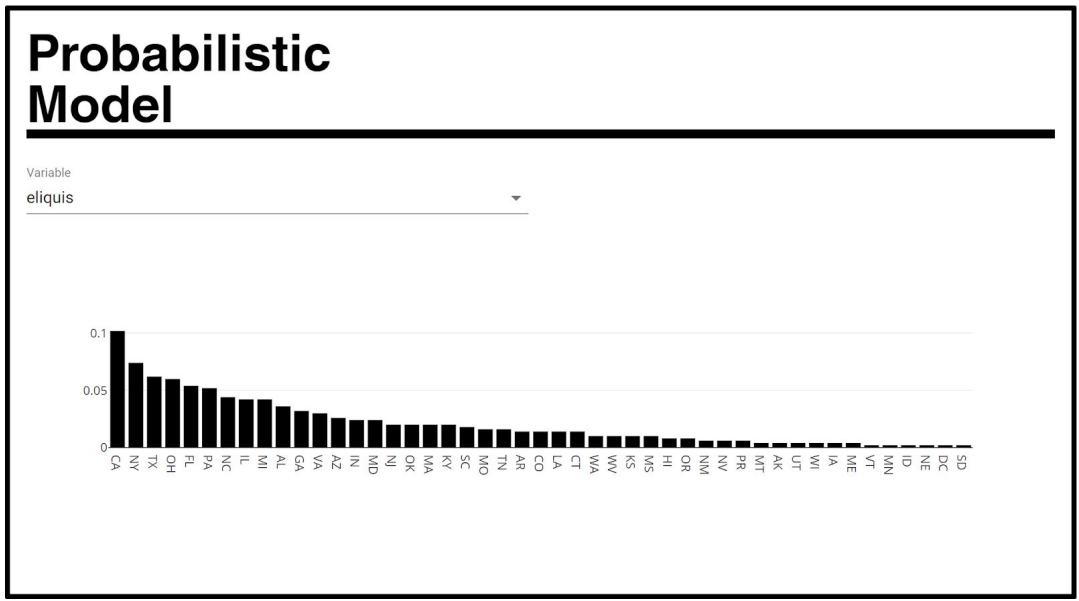

**Figure 5.** Prototype Label demonstrating the Probabilistic Model module and showcasing a hypothetical distribution for payments made towards the drug "Eliquis" across different states.

It is unavoidable that datasets collected from the real-world have relationships to demographics that the data specialist or other entities do not wish to propagate into the learned model and the inferences produced from it. For example, is a variable or an aggregate of a variable strongly correlated with the Hispanic population in a given region? To surface relationships like this, it is often necessary to explicitly compute a comparison between the dataset and demographic "ground truth" data, which is a task that can be both time consuming and challenging. The Ground Truth Correlation module (**Figure 6**) provides the data specialist initial evidence as to whether such relationships are likely, thus warranting further analysis. In order to surface any anomalies in the demographic distribution of these variables, we selected the 2010 US Census data as "ground truth" for zip code and race. The module then correlates zip code Census data with the dataset and calculates the Pearson correlation between demographics and field aggregates. To demonstrate its utility, the Label (**Figure 6, top**) highlights the negative correlations between the (sum of the) amount of payment field and demographics. A second example (**Figure 6, bottom**), highlights the positive correlation between a "spend_per_person" aggregate and demographics. This module demonstrates, in a straightforward way, specific anomalous relationships in the data that the data specialist should pay attention to during model training. In the prototype, we observe a slight positive correlation between white zip codes and payments, and a slight negative correlation between rural zip codes and payments. Toggling to per person spend underscores similar overall trends.

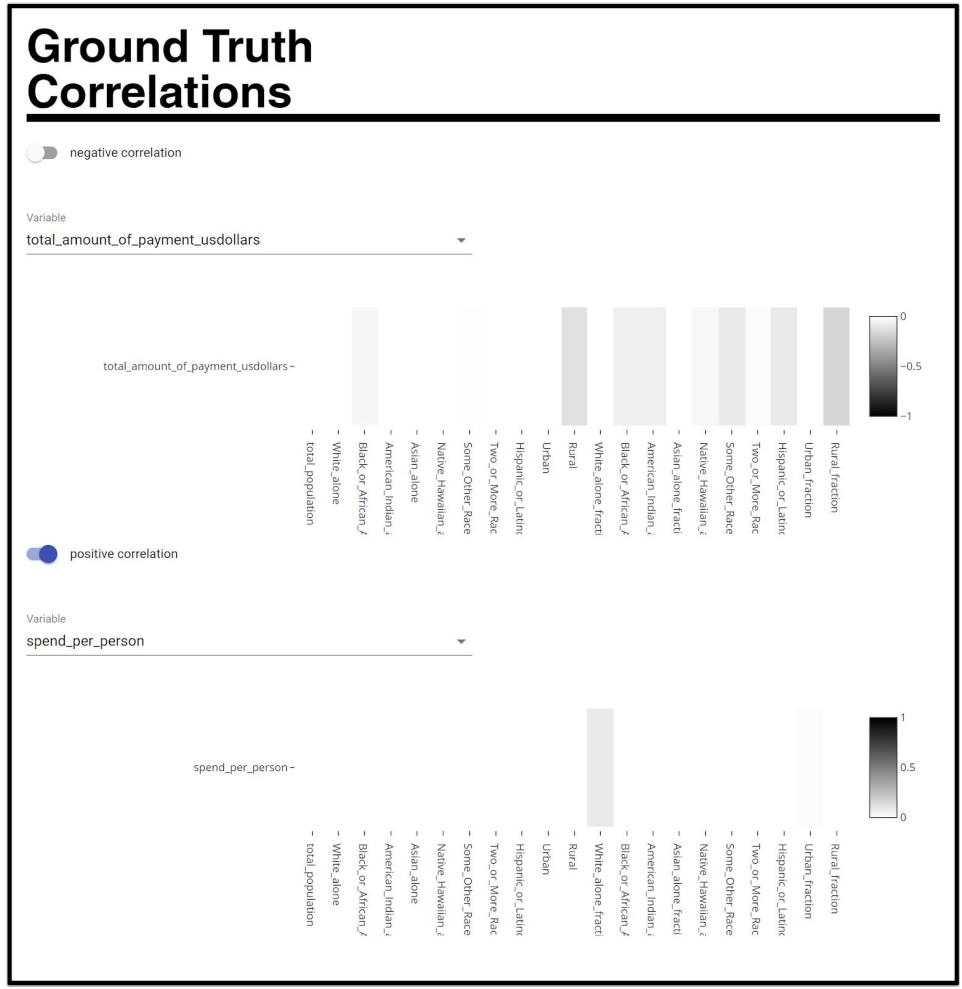

**Figure 6.** The negative (top) and positive (bottom) correlations to demographics produced by the Ground Truth Correlations module.

### 4    DISCUSSION

The Label offers many benefits. Overall, it prompts critical questions and interrogation in the preprocessing phase of model development. It also expedites decision making, which saves time in the overall model development phase without sacrificing the quality or thoroughness of the data interrogation itself, perhaps encouraging better practices at scale. These benefits apply across the spectrum of data specialists' skill and experience, but are particularly useful for those new to the field or less attuned to concerns around bias and algorithmic accountability. First, the Label creates a pre-generated "floor" for basic data interrogation in the data selection phase. It also indicates key dataset attributes in a standardized format. This gives data specialists a distilled yet comprehensive overview of the "ingredients" of the dataset, which allows for a quick and effective comparison of multiple datasets before committing to one for further investigation. It also enables the data specialist to better understand and ascertain the fitness of a dataset by scanning missing values, summary statistics of the data, correlations or proxies, and other important factors. As a result, the data specialist may discard a problematic dataset or work to improve its viability prior to utilizing it.

Improved dataset selection affords a secondary benefit: higher quality models. The Label provides data specialists improved means by which to interrogate the selected dataset during model development, previously a costly and onerous enterprise. The Ground Truth Correlation module, in particular, provides a helpful point of reference for the data specialist before model completion, and surfaces issues such as surprising variable correlations, missing data, anomalous data distributions, or other factors that could reinforce or perpetuate bias in the dataset. Addressing these factors in the model creation and training phase saves costs, time, and effort, and also could prevent bad outcomes early on, rather than addressing them after the fact.

The Label is built with scalability in mind, and with an eye towards standardization. The modular framework provides flexibility for dataset authors and publishers to identify the "right" kind and amount of information to include in a Label; over time, this could become a set of domain-specific best practices. The interactivity of the Label also permits flexibility, as insights about the dataset may arise over time. For example, the ground truth data used for comparison could evolve, rendering a previously unsuitable dataset suitable. Interactive Labels also give data specialists the ability to dive further into anomalous data, rather than simply accepting static information provided by the Label author. With some modules more subjective in nature, and with a range of domain expertise across data specialists, this is particularly important. For advanced data specialists, the flexible Label backend makes it easy to "plug-in" more complex engines. Such complex backends can provide different statistical tools; for example, the Probabilistic Computing module makes it possible to investigate low frequency variables by generating synthetic data. Synthetic data gives data specialists the ability to address incomplete data, and opens a potential path to privacy preserving data usage [34].

Lastly, the Label functioning as a proxy for the dataset itself is an intriguing, even if distant, possibility. Increased calls for accountability of AI systems demand investigation of datasets used in training models, but disclosing those datasets, even to a limited audience, may pose risks to privacy, security, and intellectual property that calls this approach into question. If the Label is able to convey the information essential for accountability in the dataset without disclosing the data itself, it would provide a valuable and much needed auditing tool [35] for AI systems while still preserving privacy, security, and proprietary information.

## 4.1   LIMITATIONS AND MITIGATIONS

There are challenges to our approach. The extensive variety of datasets used to build models raises important questions around whether the Label can generalize across data and dataset type, size, composition, and in different domains, and furthermore, whether a data specialist or domain expert will need to be involved in the creation of a Label across these different datasets. This could arise in an instance where important semantic information is atypically labeled and would be challenging to interpret automatically, such as if the field for zip code in a dataset had a custom field for "geographic area." A data specialist or domain expert may also be required when building a Label for sensitive or proprietary data, which may be accessible only to those who built the dataset and not accessible to the public. Building the Label as a modular system somewhat mitigates the complication of requiring input from a domain expert, as the framework can adapt to domain-specific best practices, and can easily support the generation of different types of Labels based on access. Within the Provenance module, it may be necessary or helpful to surface who made the Label, and what relationship they have to the dataset.

The veracity and usefulness of the Ground Truth Comparison module depends on the accuracy of the "Ground Truth" dataset, which serves as a benchmark standard and is considered objective, accurate, and provable, and with clear provenance. However, problematic ground truth data may lead to futile or even harmful comparisons. Without a realistic way to eliminate bias in all datasets, a mitigating step is to build Labels for ground truth datasets themselves. If these Labels include community feedback and comment modules, dataset authors can address the issues directly.

Further investigation is necessary to understand the feasibility and desirability of using the Label as a proxy for proprietary datasets. This would likely require that the dataset creator or controller create the Label. Another challenge is that the Label might not prompt the right questions or reactions for the data specialist, leaving certain biased data undetected. Analyses of machine bias indicate that zip codes often proxy for race, but many others proxies still exist, especially as the models themselves approach levels of complexity that are difficult or impossible for humans to comprehend and new or unexpected proxies emerge. Integrating new methods or tools to help identify proxies will be important to the industry, and our hope is that the Label will be flexible in such a way that these tools can be leveraged to create additional modules as they become available.

Finally, design of the label itself will require additional attention to determine the appropriate amount of information for presentation, comprehension, and adoption. As Kelley et al. made clear in their work on Privacy Nutrition Label [22], design is a key element in the efficacy of the label. It is worth investigating and testing the most effective presentation to drive adoption.

## 5    FUTURE DIRECTIONS

This paper and prototype are the first step toward bringing together a wide range of data specialists, from those who are creating and publishing datasets to those utilizing datasets to build models, in order to improve the quality of datasets used in AI-generated models.

Deeper research and iteration will be necessary as we continue to build additional prototypes of the Label. Creating a "nutrition label" for datasets is nascent and requires additional investigations about what information (in the form of modules or otherwise) is useful and practical to include. Based on the relatively small reach of our survey, we also recommend that a more rigorous survey be conducted to more accurately identify needs, as the survey we administered was limited in its reach, and disproportionately indexed to American and European respondents working in the private sector. The information pertinent to a data specialist will also shift based on the domain of the data, necessitating the building of additional prototypes for different kinds of datasets. The opportunities afforded by complex machine learning tools such as BayesDB in the creation of additional modules deserve more fulsome exploration to maximize the usability and usefulness of the Label.

Through building relationships with dataset publishers and circulating the Label, we hope to identify not only additional datasets for prototypes, but also to launch our Label on open datasets so that we can study the impact of the Label on the use of and conversation around the data. We will consider collaborations with colleagues from industry and academia to further drive this work, building knowledge around the impediments to adoption and considering ways that regulatory frameworks could further support the creation of a best practice or standard.

In terms of the Label ecosystem, the existence of a label for any given dataset could be notated using a mark or symbol, such as the "Conformité Européene" (CE) mark used by the European Union [36], on the author's or dataset host's webpage. Clicking on the mark would then navigate to the label viewer application and fetch the corresponding Label from a central repository where all Labels are hosted. Such a centralized archive of Labels would allow for generating usage statistics, least and most used modules, and eventually help inform future Label iterations. More importantly, a repository of this sort could act as an index of datasets without hosting the datasets themselves. For instance, API calls to such a repository could help locate datasets with queries like "MIT license dataset for facial recognition with >100k samples."

Beyond its utility as a tool, the Label could also drive a change in norms. Through using the Label, data specialists will build a habit around questioning datasets through analysis and interrogation techniques, even if a particular dataset does not include a Label. In time, the Label will facilitate an environment that encourages a broad spectrum of dataset creators, cleaners, publishers, and users to create Labels to publish alongside their datasets. This would lead to better identification of issues with data and bias, or inappropriate data collection practices, which in turn would increase data and dataset quality overall.

Looking beyond the Label itself, there are longer term opportunities for this framework and the data science community. Decisions made around the authorship and ownership model for the Label will be critical to the overall direction of the project; who will create these Labels going forward, and who will maintain them? Will there be a single place where all labels live or from where they are all linked? Additional future directions could include: building a public consortium or governing body to consider standards across the industry; creating curriculum for those collecting and working with datasets; and further exploration of appropriate ground truth data.

## 6    CONCLUSIONS

In an effort to improve the current state of practice of data analysis, we created the Dataset Nutrition Label, a diagnostic framework that provides a concise yet robust and standardized view of the core components of a dataset. We use the ProPublica Dollars for Docs dataset to create the Label prototype.

The Label serves as a proof of concept for several conceptual questions, beginning with the general feasibility of an extensible and diverse modular framework. It also confirms the possibility of mixing qualitative and quantitative modules that leverage different statistical and probabilistic modelling backend technologies in the same overall user experience. The Label integrates both static and interactive modules, underscoring the importance of using an interactive platform (such as a website) for the distribution of the Label itself. Together, this promises flexibility, scalability, and adaptability.

With the Label, data specialists can efficiently compare, select, and interrogate datasets. Additionally, certain modules afford the ability to check for issues with the dataset before and during model development, surface anomalies and potentially dangerous proxies, and find new insights into the data at hand. As a result, data specialists have a better, more efficient process of data interrogation, which will produce higher quality AI models. The Label is a useful, practical, timely, and necessary intervention in the development of AI models, and a first step in a broader effort toward improving the outcomes of AI systems that play an increasingly central role in our lives.

# 7   ACKNOWLEDGMENTS

We are grateful to the ProPublica team, including Celeste LeCompte, Ryann Jones, Scott Klein, and Hannah Fresques, for their generosity in providing the Dollars for Docs dataset and for their assistance throughout prototype development, and to the BayesDB team in the Probabilistic Computing Group at MIT, including Vikash Mansinghka, Sara Rendtorff-Smith, and Ulrich Schaechtle for their valuable work and ongoing advice and assistance. We are also thankful to Patrick Gage Kelley for bringing key work to our attention and for his constructive feedback, and to the 2018 Assembly Cohort and Advisory Board, in particular Matt Taylor, Jack Clark, Rachel Kalmar, Kathy Pham, James Mickens, Andy Ellis, and Nathan Freitas; the City of Boston Office of New Urban Mechanics; and Eric Breck and Mahima Pushkarna of Google Brain for productive and insightful discussions. This work was made possible by the Assembly program led by Jonathan Zittrain of the Berkman Klein Center for Internet & Society and Joi Ito of the MIT Media Lab.

## Dataset Facts
ProPublica's Dollars for Docs Data

### Metadata
| | |
|---|---|
| Filename | 201612v1-docdollars-product_payments |
| Format | csv |
| Url | https://projects.propublica.org/docdollars/ |
| Domain | healthcare |
| Keywords | Physicians, drugs, medicine, pharmaceutical, transactions |
| Type | tabular |
| Rows | 500 |
| Columns | 18 |
| Missing | 5.2% |
| License | cc |
| Released | JAN 2017 |
| Range | |
| From | AUG 2013 |
| To | DEC 2015 |
| Description | This is the data used in ProPublica's Dollars for Docs news application. It is primarily based on CMS's Open Payments data, but we have added a few features. ProPublica has standardized drug, device and manufacturer names, and made a flattened table (product_payments) that allows for easier aggregating payments associated with each drug/device. In [1], one payment record can be attributed to up to five different drugs or medical devices. This table flattens the payments out so that each drug/device related to each payment gets its own line. |

### Provenance
**Source**
| | |
|---|---|
| Name | U.S. Centers for Medicare & Medicaid Services |
| Url | https://www.cms.gov/OpenPayments/ |
| Email | openpayments@cms.hhs.gov |

**Author**
| | |
|---|---|
| Name | Propublica |
| Url | https://www.propublica.org/datastore/ |
| Email | data.store@propublica.org |

### Variables
| | |
|---|---|
| Id | A unique ID number for this payment & product combination. This is assigned by ProPublica for internal use |
| Applicable_manufacturer_or_applicable_gpo_making_payment_id | ID of the applicable manufacturer or submitting applicable GPO making the payment or other transfer of value |
| Date_of_payment | If a singular payment, then this is the actual date the payment was issued; if a series of payments or an aggregated set of payments, this is the date of the first payment to the covered recipient in this program year |
| General_transaction_id | System-assigned identifier to the general transaction at the time of submission |
| Program_year | The calendar year for which the payment is reported in Open Payments |
| Product_name | Derived from the 'name_of_associated_covered_drug_or_biologicalX' field (for drugs) or 'name_of_associated_covered_device_or_medical_supplyX' field (for medical devices). Where possible, multiple versions of the same product are converted to the same product_name (i.e. records for 'Zorvolex 65mg' and 'Zorvolex 35mg' will be converted to 'Zorvolex'). The original value is contained in original_product_name |
| Original_product_name | A copy of the original name_of_associated_covered_drug_or_biologicalX' field (for drugs) or 'name_of_associated_covered_device_or_medical_supplyX' field (for medical devices) |
| Product_ndc | If the product is a drug, this a copy of the original 'ndc_of_associated_covered_drug_or_biologicalX' field |
| Product_is_drug | 't' if the product is a drug (contained in a 'name_of_associated_covered_drug_or_biologicalX' field). 'f' if the product is a medical device (contained in a 'name_of_associated_covered_device_or_medical_supplyX' field) |
| Payment_has_many | 't' if the original payment record included data on more than one drug or device, i.e. 'name_of_associated_covered_drug_or_biological1' and 'name_of_associated_covered_drug_or_biological2', 'name_of_associated_covered_device_or_medical_supply1' and 'name_of_associated_covered_device_or_medical_supply2', etc. |
| Teaching_hospital_id | Open Payments system-generated unique identifier of the teaching hospital receiving the payment or other transfer of value |
| Physician_profile_id | ID of the physician receiving the payment or other transfer of value |
| Recipient_state | The state or territory abbreviation of the primary business address of the physician or teaching hospital or non-covered recipient entity receiving the payment or other transfer of value if the primary business address is in the United States |
| Applicable_manufacturer_or_applicable_gpo_making_payment_name | Textual proper name of the applicable manufacturer or applicable GPO making the payment or other transfer of value. This field has been standardized to eliminate different names attributable solely to punctuation |
| Teaching_hospital_ccn | A unique identifying number (CMS Certification Number) of the Teaching Hospital receiving the payment or other transfer of value |
| Product_slug | Used internally at ProPublica for web display on the Dollars for Docs app. You can pull up the corresponding Dollars for Docs page for a product by appending product_slug to https://projects.propublica.org/docdollars/products/, i.e. https://projects.propublica.org/docdollars/products/device-dental-cabinetry |
| Total_amount_of_payment_usdollars | U.S. dollar amount of payment or other transfer of value to recipient (manufacturer must convert to dollar currency if necessary) |
| Number_of_payments_included_in_total_amount | The number of discrete payments being reported in the 'Total Amount of Payment' data element |

**Supplement figure 1.** Prototype Label demonstrating the metadata, provenance, and variables modules.

## Statistics

### Ordinal

| name | type | count | uniqueEntries | mostFrequent | leastFrequent | missing |
|---|---|---|---|---|---|---|
| id | number | 500 | 488 including mi… | missing value (13) | multiple detected | 2.60% |
| applicable_man… | number | 500 | 4 | 100000000232 (… | multiple detected | 0% |
| date_of_payment | date | 500 | 213 including mi… | missing value (27) | multiple detected | 5.40% |
| general_transac… | number | 500 | 467 including mi… | missing value (34) | multiple detected | 6.80% |
| program_year | number | 500 | 2 including missi… | 2014 (495) | missing value (5) | 1.00% |

### Nominal

| name | type | count | uniqueEntries | mostFrequent | leastFrequent | missing |
|---|---|---|---|---|---|---|
| product_name | string | 500 | 16 including mis… | Xarelto (200) | Aciphex (1) | 3.20% |
| original_product… | string | 500 | 15 | Xarelto (212) | Aciphex (1) | 0% |
| product_ndc | number | 500 | 21 including mis… | 5045857810 (201) | multiple detected | 5.00% |
| product_is_drug | boolean | 500 | 2 including missi… | t (492) | missing value (8) | 1.60% |
| payment_has_m… | boolean | 500 | 3 including missi… | f (267) | missing value (29) | 5.80% |
| teaching_hospit… | number | 500 | 2 including missi… | 0 (464) | missing value (36) | 7.20% |
| physician_profile… | number | 500 | 230 including mi… | missing value (32) | multiple detected | 6.40% |
| recipient_state | string | 500 | 40 | CA (56) | multiple detected | 0% |
| applicable_man… | string | 500 | 5 including missi… | Janssen Pharm… | multiple detected | 7.00% |
| teaching_hospit… | number | 500 | 2 including missi… | 0 (481) | missing value (19) | 3.80% |
| product_slug | string | 500 | 15 including mis… | drug-xarelto (196) | drug-aciphex (1) | 8.20% |

### Continuous

| name | type | count | min | median | max | mean | standardD… | missing | zeros |
|---|---|---|---|---|---|---|---|---|---|
| total_amo… | number | 500 | 0.14 | 14.00 | 5000 | 134.21 | 501.99 | 9.40% | 0% |

### Discrete

| name | type | count | min | median | max | mean | standardD… | missing | zeros |
|---|---|---|---|---|---|---|---|---|---|
| number_o… | number | 500 | 1 | 1.00 | 1 | 1.00 | 0.00 | 4.80% | 0% |

**Supplement figure 2.** Prototype Label demonstrating the Statistics module, splitting the variables into 4 groups: ordinal, nominal, continuous, and discrete.